\documentclass[12pt]{iopart}
\def\0{\mbox{\tiny $0$}}
\def\1{\mbox{\tiny $1$}}
\def\2{\mbox{\tiny $2$}}
\def\3{\mbox{\tiny $3$}}
\def\4{\mbox{\tiny $4$}}
\def\5{\mbox{\tiny $5$}}
\def\6{\mbox{\tiny $6$}}
\def\7{\mbox{\tiny $7$}}
\def\8{\mbox{\tiny $8$}}
\def\9{\mbox{\tiny $9$}}

\def\f14{\mbox{\tiny $\frac{1}{4}$}}

\begin{document}

\title[Gauge boson masses for a minimal $SU(4)_{EW} \otimes U(1)_{B-l}$ model]{The gauge boson masses for a minimal $SU(4)_{EW} \otimes U(1)_{B-l}$ model for electroweak interactions with left-right symmetry}

\author{A. E. Bernardini\dag\ 
\footnote[3]{alexeb@ifi.unicamp.br}
}

\address{\dag\ Instituto de Física Gleb Wataghin, UNICAMP,\\
PO Box 6165, SP 13083-970, Campinas, SP, Brazil}

\date{\today}

\begin{abstract}
Following the general procedure of spontaneous symmetry breaking of a $SU(4)_{EW} \otimes U(1)_{B-l}$ electroweak gauge group with left-right symmetry, we obtain the gauge boson masses and currents for the minimal version of the model.
The physical eigenstates for neutral gauge bosons are determined by introducing two mixing angles $\theta_4$ and $\theta_3$ which are related to the electroweak mixing angle $\theta_w$ at the unification scale.
By introducing some physical approaches in order to simplify the calculations, we calculate the charged and neutral currents.
Differently from other previous propositions, the results are obtained from a theoretical constraint upon the coupling constants as a consequence of embedding the symmetry into the Pati-Salam electroweak coupling $SU(4)_{EW} \otimes SU(4)_{PS}$.
\end{abstract}




\maketitle 

\section{\label{sec1}Introduction}

We know that the standard model (SM) \cite{G61,S68,W67}
based on the gauge symmetry $SU(3)_{c} \otimes SU(2)_L\otimes U(1)_Y$,
in spite of not being the ultimate theory, describes most of the observed properties of charged leptons and quarks. 
Each one of the fermion generations of the SM is anomaly-free, which is true
for many extensions of it, including the popular grand unification theories (GUTs) \cite{GG74}.
At the moment, the necessity to go beyond it, from the experimental point of view,
comes from leptogenesis and certain intriguing features
of quark-lepton masses and mixing \cite{Pat03} 
as well as from neutrino oscillation data \cite{Kay04} that clearly indicate the massiveness of neutrinos.

In this context, a rich discussion about the hierarchy problem \cite{AHD98} has emerged
and an interesting class of models has been proposed \cite{PP95,PT93,M99,F92,F98,FF99}. 
In some of these models, each generation has anomalies which are canceled when the
number of generations is chosen to be three \cite{S80}.
At the same time, another class of models has considered the existence of a
kind of right-handed neutrino coupling so that an anomaly free model can be constructed.
In this category of models, the boson sector analysis can provide a
series of phenomenological constraints which brings up additional information for
GUTs.
In particular, Pleitez {\it et al.} have proposed the
341 \cite{PP95} and the 331 \cite{PT93} models
where they have described an electroweak unification structure
using respectively the gauge groups $SU(4)_L$ and $SU(3)_L$.
In both of the models the chiral right-handed neutrinos are
not considered and some of the multiplets of the group allow
the existence of quarks and leptons with {\em exotic} charges. 
Long and Pal proposed a variation of Pleitez 331 model
\cite{LP98} where the right-handed neutrinos appear in the same
multiplet representation of some new
{\em exotic} fermions.

Our purpose concerns with the analysis of an electroweak gauge model based on the group
$SU(4)_{EW} \otimes U(1)_{B-l}$\footnote{The subindex $EW$ denotes an ``electroweak'' symmetry and $B-l$
corresponds to the barion minus lepton number.} dealing with left and right-handed ($L$ and $R$) chiral fermions.
Due to the left-right symmetry, differently from the above quoted models, it is not necessary
to consider new {\em exotic} fermionic particles to construct an anomaly free model.
The $SU(4)_{EW} \otimes U(1)_{B-l}$ results from the spontaneous symmetry breaking (SSB)
of the $SU(4)_{PS} \otimes SU(4)_{EW}$ where the $SU(4)_{PS}$ symmetry is inspired
by the Pati-Salam idea of quark-lepton unification \cite{PS74} with the lepton number being
described as a fourth color number,
\small\begin{equation}
\begin{array}{lcccc}
SU(4)_{PS} & \otimes &  SU(4)_{EW}&&  \\
   g_{4}     &         &  g_{4} &&   \\
   &\Downarrow&  &&            \\
   &\phi_{15-dim}&  &&            \\
 SU(3)_{c}& \otimes &  SU(4)_{EW} &\otimes& U(1)_{B-l}\\
    g_{c}     &         &  g_{4}  &&g_{Bl}  \\
\end{array}
\label{eee1}.
\end{equation}\normalsize
In order to find a complete fermion representation, it is natural
to include all fermions of the first generation of the model in the following matrix representation,
\small\begin{equation}
\Psi = \left[
\begin{array}{llll}
u_{1L}  & u_{2L}  & u_{3L}  & \nu^e_{L}  \\
d_{1L}  & d_{2L}  & d_{3L}  & e_{L}      \\
u_{1R}  & u_{2R}  & u_{3R}  & \nu^e_{R}  \\
d_{1R}  & d_{2R}  & d_{3R}  & e_{R}
\end{array}
\right].
\label{bbb3}
\end{equation}\normalsize
At the same time, we consider the existence of a universal coupling constant $g_G$ in a way that we can establish
\small\begin{equation}
G \supset SU(4)_{PS}  \otimes   SU(4)_{EW}
\label{bbb3b}.
\end{equation}\normalsize 
Thus, the free Lagrangian can be described by
\small\begin{equation}
\mathcal{L} = Tr \left[\overline{\Psi} \gamma^{\mu} \mbox{\boldmath$D_{\mu}$} \Psi \right] + \mbox{coupling and interactions}
\label{bbb3c},
\end{equation}\normalsize
with
\small\begin{equation}
\mbox{\boldmath$D_{\mu}$} \Psi =  \partial _{\mu} \Psi - i g_G \left( h^a_{\mu (EW)} H^a \Psi + h^b_{\mu (PS)} \Psi H^b \right),
\label{bbb31}
\end{equation}\normalsize
where $H^{a}$ and $H^{b}$ are the $SU(4)$ generators for, respectively, $EW$ and $PS$ interactions,
which are not exactly in the same irreducible representations.
By observing the the $EW$ sector of the 44 model, the sequence of SSB can be described by one of the following chains
\small\begin{eqnarray}
&&
\begin{array}{cccc}
\mbox{a)}&SU(4)_{EW}   & \otimes & U(1)_{B-l} \\
  & g_{4}     &         &  g_{Bl}    \\
  & &\Downarrow&             \\
  & &\chi_{4-dim}&             \\
&SU(3)_{L(R)} & \otimes & U(1)_{Y_{SU(3)}}  \nonumber\\
   &g_{3}     &         &  g_{Y_{SU(3)}}\\
\end{array}
\label{eee2},\\
&&\nonumber\\
&&
\begin{array}{cccccc}
\mbox{b)}&SU(4)_{EW}   & \otimes & U(1)_{B-l} \\
  & g_{4}     &         &  g_{Bl}    \\
  & &\Downarrow&             \\
  & &\Phi_{15-dim}&            \\
&SU(2)_{L}    & \otimes & SU(2)_{R} & \otimes &  U(1)_{B-l}  (\otimes  U(1)_{chiral})    \nonumber\\
 &  g_{L}     &         &     g_{R} &         &    g_{Bl}    \\
\end{array}
\label{bbb2b},\\
&&\nonumber\\
&&
\begin{array}{cccccc}
\mbox{c)}&SU(4)_{EW}   & \otimes & U(1)_{B-l} &&\\
  & g_{4}     &         &  g_{Bl}    \\
    & &\Downarrow&             \\
	& &\Phi_{15-dim}&             \\
&SU(2)_{L}    & \otimes & U(1)_Y (\otimes  U(1)_{chiral})     \nonumber\\
  & g         &         &  g^{\prime} \\
\end{array}
\label{bbb2c}.
\end{eqnarray}\normalsize
The way like the symmetries are broken depends on the number and on the multiplet structure of the Higgs bosons included in the model.
By following the SSB chains $a)$, $b)$ or $c)$, we can independently obtain the same gauge boson masses and the
coupling currents which are different from those obtained in a 422 model \cite{F98,FF99}. 
Therefore, we aim to investigate the gauge boson properties of the $EW$ sector of this model 
by restricting our calculations to the SSB chain from expression a) in the following way
\small\begin{equation}
\begin{array}{ccc}
SU(4)_{EW}   & \otimes & U(1)_{B-l} \\
   g_{4}     &         &  g_{Bl}    \\
   &\Downarrow&             \\
     &\chi_{4}&             \\
SU(3)_{L(R)} & \otimes & U(1)_{Y_{SU(3)}}  \\
   g_{3}     &         &  g_{Y_{SU(3)}}    \\
 &\Downarrow&             \\
     &\chi_{3}&             \\
	 SU(2)_{L}    & \otimes & U(1)_Y      \\
   g         &         &  g^{\prime} \\
   &\Downarrow&             \\
     &\rho,   \eta&                  \\
U(1)_{EM}    &         &  \\
  e        &         &  \\
\end{array}
\label{eee2a}.
\end{equation}\normalsize
where all the scalar {\em Higgs} bosons are in a $4-dim$ representation.

In order to organize the ideas which are discussed in the manuscript, we have stated the following division.
In section II, we recall some features of $SU(4)_{EW} \otimes U(1)_{B-l}$ and we
introduce the study of the gauge boson sector.
Some constraints on the gauge boson mixing and masses are obtained in section III.
The charged and neutral currents are obtained in section IV.
In section V,
we compare some phenomenological possibilities of studying decay rates and cross sections at unification energy scales in order to give directions to a more suitable phenomenological analysis.
Finally, our conclusions are summarized in section VI.

\section{\label{sec2}The model and the gauge bosons}

The model is constructed with basis on the gauge group described in (\ref{eee1}) and,
by simplicity, we will not consider the color $SU(3)_{c}$ coupling implications on the
subsequent results.
It takes into account the number of quarks and leptons already known and it
includes chiral right-handed neutrinos which were not (yet) experimentally detected.
The fermions of the first generation of the model are the leptons
$\nu_L$, $e_L$, $\nu_R$ and $e_R$ and the quarks $u_L$, $d_L$, $u_R$ and $d_R$,
which follow a similar arrangement for the next generations.
Under the gauge symmetry (\ref{eee2a}), the three fermion generations transform as \footnote{If we permute the third
and fourth lines, the mixing and gauge boson mass results will not be modified.}
\small\begin{equation}
\psi^{\ell}_{(LR)}    = \left(\begin{array}{l} \nu^e_{L}\\ e_L    \\ \nu^e_{R}\\ e_R   \end{array}\right) \sim (\mathbf{4}, - 1)
\label{eee3}.
\end{equation}\normalsize
\small\begin{equation}
\psi^{q}_{(LR)} = \left(\begin{array}{l} u_{L}   \\ d_{L}  \\ u_{R}   \\ d_{R} \end{array}\right) \sim (\mathbf{4}, +\frac{1}{3})
\label{eee33}
\end{equation}\normalsize
where $q$ is the color index.
In both cases, each generation transforms identically and we
can easily verify the gauge anomaly cancellation when the above choice of gauge quantum numbers is adopted.
In spite of placing different Lorentz objects in the same gauge multiplets the theory is still Lorentz invariant
since the interaction Lagrangian is kept invariant\footnote{The covariance of the Lagrangian prohibits Lorentz violating gauge couplings between different chiral objects.
Such couplings naturally (mathematically) disappear.
In a similar way, before the SSB, one can easily observe that any Lagrangian chiral conversion (Lorentz violating) term disappears.
The generalized Yukawa coupling given by $\overline{\psi} \Gamma^i H^a \psi \phi^a$ for which the Lorentz invariance can be demonstrated (when $\phi$ is in the $15$-dim adjoint representation with $a =\,1,\,2,\,...,\,15$),
can be {\em decomposed} into:
\begin{eqnarray}
\fl\frac{1}{2} \{ (\overline{\ell_L} \nu_L + \overline{\nu_L} \ell_L) \phi_1
                +    i(\overline{\ell_L} \nu_L - \overline{\nu_L} \ell_L) \phi_2    
				+     (\overline{\nu_L} \nu_L   - \overline{\ell_L} \ell_L)\phi_3
				+     (\overline{\nu_L} \nu_R  +  \overline{\nu_R} \nu_L) \phi_4\nonumber\\ 
		       \lo+   i(\overline{\nu_R} \nu_L   - \overline{\nu_L} \nu_R)  \phi_5                                                                     
			    +     (\overline{\ell_L} \nu_R + \overline{\nu_R} \ell_L)\phi_6     
		        +    i(\overline{\nu_R} \ell_L  - \overline{\ell_L} \nu_R)\phi_7     
			    +     (\overline{\nu_R} \nu_R   - \overline{\ell_R} \ell_R)  \phi_8\nonumber\\
		       \lo+    (\overline{\ell_R} \nu_R + \overline{\nu_R} \ell_R) \phi_9   
		        +    i(\overline{\ell_R} \nu_R  - \overline{\nu_R} \ell_R)   \phi_{10} 
		        +     (\overline{\ell_R} \ell_L + \overline{\ell_L} \ell_R) \phi_{11}
	            +    i(\overline{\ell_R} \ell_L - \overline{\ell_L} \ell_R)  \phi_{12}\nonumber\\
	           \lo+   (\overline{\ell_R} \nu_L  +  \overline{\nu_L} \ell_R) \phi_{13}  
				+    i(\overline{\ell_R} \nu_L   - \overline{\nu_L} \ell_R) \phi_{14} 
				+   \frac{1}{\sqrt{2}}(\overline{\nu_L} \nu_L   
				+ \overline{\ell_L} \ell_L - \overline{\nu_R} \nu_R   
				- \overline{\ell_R} \ell_R)  \phi_{15}\},\nonumber
\end{eqnarray}
from which we obtain just eight non-vanishing terms:
\begin{eqnarray}
   \fl  (\overline{\nu_L} \nu_R  +  \overline{\nu_R} \nu_L) \phi_4      +   i(\overline{\nu_R} \nu_L   - \overline{\nu_L} \nu_R)    \phi_5
   +   (\overline{\ell_L} \nu_R + \overline{\nu_R} \ell_L) \phi_6      +   i(\overline{\nu_R} \ell_L  - \overline{\ell_L} \nu_R)   \phi_7    \nonumber \\
   \lo+ (\overline{\ell_R} \ell_L + \overline{\ell_L} \ell_R) \phi_{11} +   i(\overline{\ell_R} \ell_L - \overline{\ell_L} \ell_R)  \phi_{12}
   +   (\overline{\ell_R} \nu_L  +  \overline{\nu_L} \ell_R) \phi_{13} +   i(\overline{\ell_R} \nu_L   - \overline{\nu_L} \ell_R) \phi_{14}  \nonumber \\
   \fl= (\overline{\ell_L} \nu_R) (\phi_4 - i \phi_5)       +    (\overline{\nu_R} \ell_L)  (\phi_4 + i \phi_5)      
   +   (\overline{\nu_L} \nu_R ) (\phi_6 - i \phi_7)       +    (\overline{\nu_R} \nu_L)   (\phi_6 + i \phi_7)      \nonumber \\
   \lo+ (\overline{\ell_L} \ell_R (\phi_{11} - i \phi_{12}) +    (\overline{\ell_R} \ell_L) (\phi_{11} + i\phi_{12}) 
   +   (\overline{\nu_L} \ell_R) (\phi_{13} - i \phi_{14}) +    (\overline{\ell_R} \nu_L)  (\phi_{13} + i \phi_{14}).\nonumber
\end{eqnarray}
which disappear by computing the VEV of $\phi _{15 - dim}$ (all the above remnant chiral conversion terms disappear).}
The symmetry breaking could be achieved with three or four $SU(4)_{EW}$ scalar $4-dim$ multiplets,
but the last choice provides an additional free parameter which can be essential in determining the fermion masses.
In this case, the respective vacuum expectation values (VEV) of the scalar bosons $\chi_4$, $\chi_3$, $\rho$ and $\eta$, are described by
\small\begin{equation}
\chi_4 =  \frac{1}{\sqrt{2}}\left(\begin{array}{c} \chi_{14}^{+}\\ \chi_{24}^{0}\\ \chi_{34}^{+}\\ \chi_{44}^0 \end{array}\right)\sim (\mathbf{4}, + 1)
\label{eee4a1},
\end{equation}\normalsize
\small\begin{equation}
\langle\chi_4\rangle^T = \chi_{o4}^T = \frac{1}{\sqrt{2}}\left(0,~ 0, ~ 0,~ w\right)
\label{eee4a2},
\end{equation}\normalsize
\small\begin{equation}
\chi_3 =  \frac{1}{\sqrt{2}}\left(\begin{array}{c} \chi_{13}^{0}\\ \chi_{23}^{-}\\ \chi_{33}^{0}\\ \chi_{43}^{-} \end{array}\right)\sim (\mathbf{4}, - 1)
\label{eee4b1},
\end{equation}\normalsize
\small\begin{equation}
\langle\chi_3\rangle^T = \chi_{o3}^T = \frac{1}{\sqrt{2}}\left(0,~ 0,~ d,~ 0\right)
\label{eee4b2},
\end{equation}\normalsize
\small\begin{equation}
\rho = \frac{1}{\sqrt{2}}\left(\begin{array}{c} \rho_{1}^{+}\\ \rho_{2}^0 \\ \rho_{3}^{+} \\ \rho_{4}^{0} \end{array}\right)\sim (\mathbf{4}, + 1)
\label{eee4c1},
\end{equation}\normalsize
\small\begin{equation}
\langle\rho\rangle^T = \rho_{o}^T = \frac{1}{\sqrt{2}}\left(0,~ v,~ 0, ~ 0\right)
\label{eee4c2},
\end{equation}\normalsize
\small\begin{equation}
\eta = \frac{1}{\sqrt{2}}\left(\begin{array}{c} \eta_{1}^{0} \\ \eta_{2}^{-} \\ \eta_{3}^{0} \\ \eta_{4}^{-} \end{array}\right)\sim (\mathbf{4}, - 1)
\label{eee4d1},
\end{equation}\normalsize
\small\begin{equation}
\langle\eta\rangle^T = \eta_{o}^T = \frac{1}{\sqrt{2}}\left(u,~ 0 ,~ 0, ~ 0\right)
\label{eee4d2}.
\end{equation}\normalsize
The $B-l$ quantum numbers given by the upper index are chosen to maintain the invariance under $U(1)_{EM}$ transformations.
when the SSB chain for the SM gauge symmetry is given by (\ref{eee2}). 
Consequently, the electric charge and the correspondent hypercharges are respectively defined as
\small\begin{equation}
Q = \sqrt{\frac{2}{3}} H_{15} - \frac{1}{\sqrt{3}} H_8 + H_3 + \frac{(B-l)}{2}
\label{eee6},
\end{equation}\normalsize
and
\small\begin{equation}
Y_{SU_3} = \sqrt{\frac{8}{3}} H_{15} + (B-l)
\label{eee7},
\end{equation}\normalsize
\small\begin{equation}
Y = - \sqrt{\frac{4}{3}} H_{8} + Y_{SU_3}
\label{eee8},
\end{equation}\normalsize
where
\small\begin{equation}
\begin{array}{lcr}
H_3     &= & \frac{1}{2}         ~diag \left[1,   -1,  ~~~0,   ~~~0\right],\\
&&\\
H_8     &= & \frac{1}{2\sqrt{3}} ~diag \left[1,~~~1,     -2,   ~~~0\right],\\
&&\\
H_{15}  &= & \frac{1}{2\sqrt{6}} ~diag \left[1,~~~1,  ~~~1,     -3\right].
\end{array}
\end{equation}\normalsize

The gauge bosons $h_\mu^a$ of our model form a multiplet $15-dim$ related to $SU(4)_{EW}$ and a singlet $d_\mu$ related to $U(1)_{B-l}$.
There are else the massless gluons $G_\mu^a$ associated with $SU(3)_c$ which decouple from the neutral gauge boson mass
matrix\footnote{However, in this simplified analysis, we will not consider the terms which contain gluons in the covariant derivative.}.
The fermion mass are achieved by means of Yukawa Lagrangians which are not being considered at
this analysis.
By the way, the gauge boson mass matrix arises from the Higgs boson kinetic term
\small\begin{eqnarray}
\Delta \mathcal{L}_{kinetic}&=& \mid D_\mu \chi_4\mid ^2 + \mid D_\mu \chi_3\mid ^2 + \mid D_\mu \rho\mid ^2 + \mid D_\mu \eta\mid ^2  \nonumber\\
							  &=&  \partial _\mu \eta_4 \partial^\mu\eta_4 + \partial _\mu \eta_3 \partial^\mu\eta_3 + \partial _\mu \eta_2 \partial^\mu\eta_2 + \partial _\mu \eta_1 \partial^\mu\eta_1 \nonumber\\ 
                              && +  \chi_4^{\prime \dagger}(g_{4} h_\mu^a H^a + \,  \frac{1}{2} g_{Bl}   d_\mu)(g_{4} h_\mu^b H^b + \,  \frac{1}{2} g_{Bl}   d_\mu)\chi_4^{\prime} \nonumber\\
							  && +  \chi_3^{\prime \dagger}(g_{4} h_\mu^a H^a - \,  \frac{1}{2} g_{Bl}   d_\mu)(g_{4} h_\mu^b H^b - \,  \frac{1}{2} g_{Bl}   d_\mu)\chi_3^{\prime} \nonumber\\
                              && +  \rho^{\prime \dagger}(g_{4} h_\mu^a H^a + \,  \frac{1}{2} g_{Bl}   d_\mu)(g_{4} h_\mu^b H^b + \,  \frac{1}{2} g_{Bl}   d_\mu)\rho^{\prime} \nonumber\\
							  && +  \eta^{\prime \dagger}(g_{4} h_\mu^a H^a - \,  \frac{1}{2} g_{Bl}   d_\mu)(g_{4} h_\mu^b H^b - \,  \frac{1}{2} g_{Bl}   d_\mu)\eta^{\prime}
\label{eee13}
\end{eqnarray}\normalsize
with $\chi_4^{\prime}$, $\chi_3^{\prime}$, $\rho^{\prime}$ and $\eta^{\prime}$ coupled to $h_\mu^a$ and $d_\mu$ in the unitary gauge.
The covariant derivative is
\small\begin{equation}
D_\mu  =   \partial _\mu  -i g_{(4)} \left(\sum_{i = 1}^{15} h^a_\mu H^a\right) - i g_{(Bl)}\frac{B -l}{2}d_{\mu}
\label{eee14},
\end{equation}\normalsize
where $H_\mu^a$ are the $SU(4)_{EW}$ generators.
The non-Hermitian gauge bosons have the following masses
\small\begin{equation}
\begin{array}{lclcccrcl}
\sqrt{2}\,W ^{\mp}_\mu    & = &  h_\mu^1     \pm i h_\mu^2    & & \Rightarrow&  & M^2_W &=& \frac{1}{4} g_4^2 \left(u^2 + v^2 \right), \\
\sqrt{2}\,U ^{ 0(\dagger)}_\mu  & = &  h_\mu^4     \pm i h_\mu^5    & & \Rightarrow&  & M^2_U &=& \frac{1}{4} g_4^2 \left(v^2 + d^2 \right), \\
\sqrt{2}\,V ^{\pm}_\mu    & = &  h_\mu^6     \pm i h_\mu^7    & & \Rightarrow&  & M^2_V &=& \frac{1}{4} g_4^2 \left(u^2 + d^2 \right), \\
\sqrt{2}\,X ^{\mp}_\mu    & = &  h_\mu^9     \pm i h_\mu^{10} & & \Rightarrow&  & M^2_X &=& \frac{1}{4} g_4^2 \left(w^2 + d^2 \right), \\
\sqrt{2}\,Y ^{ 0(\dagger)}_\mu  & = &  h_\mu^{11}  \pm i h_\mu^{12} & & \Rightarrow&  & M^2_Y &=& \frac{1}{4} g_4^2 \left(u^2 + w^2 \right),\\
\sqrt{2}\,T ^{\mp}_\mu    & = &  h_\mu^{13}  \pm i h_\mu^{14} & & \Rightarrow&  & M^2_T &=& \frac{1}{4} g_4^2 \left(v^2 + w^2 \right).
\label{eee15}
\end{array}
\end{equation}\normalsize
and the neutral Hermitian gauge bosons have the mass matrix $\mathbf{M}^2$ written in the ${\mathbf W}$ basis
\small\begin{equation}
{\mathbf W}^{\dagger} = \left(
\begin{array}{cccc}
h_\mu^3 & h_\mu^8 &  h_\mu^{15} & d_\mu
\end{array}\right),
\label{eee17}
\end{equation}\normalsize
where
\small\begin{equation}
\Delta \mathcal{L}_{mass}=\frac{1}{2} {\mathbf W}^{\dagger}\cdot {\mathbf M}^2 \cdot{\mathbf W}
\label{eee151},
\end{equation}\normalsize
\small\begin{equation}
\fl~~~~~~~~~~{\mathbf M}^2=\frac{g_4^2}{4}\left[
\begin{array}{rrrr}
v^2  + u^2                  & \frac{v^2-u^2}{\sqrt{3}}                   & \frac{v^2-u^2}{\sqrt{6}}                          & -(v^2  + u^2 ) t   \\
\frac{v^2-u^2}{\sqrt{3}}    & \frac{v^2+u^2+4d^2}{3}                     & \frac{v^2+u^2-2d^2}{3\sqrt{2}}                    & \frac{u^2  - v^2 + 2 d^2}{\sqrt{3}} t\\
\frac{v^2-u^2}{\sqrt{6}}    & \frac{v^2+u^2-2d^2 }{3\sqrt{2}}            & \frac{v^2+u^2+d^2+9w^2}{6}                        & \frac{u^2  - v^2 -  d^2 - 3w^2}{\sqrt{6}} t \\
-(v^2  + u^2) t~             & \frac{u^2  - v^2 + 2 d^2}{\sqrt{3}} t~    & \frac{u^2  - v^2 -  d^2 - 3w^2}{\sqrt{6}} t~   & (v^2  + u^2 + d^2 + w^2) t^2~
\end{array}
\right],
\label{eee16}
\end{equation}\normalsize
with $t= \frac{g_{Bl}}{g_4}$.

The mass matrix $\mathbf{M}^2$ (\ref{eee16}) has a null determinant which allow us to identify immediately
a massless photon $A_\mu$, the $U(1)_{EM}$ gauge boson, as well as the massive bosons $Z_\mu^A$, $Z_\mu^B$ and $Z_\mu^C$ for which we have set $M_{Z_A}<M_{Z_B}<M_{Z_C}$.
Meanwhile, ${\mathbf M}^2$ can also be written in a new basis ${\mathbf Z}$ as ${\mathbf{\mathcal M}}^2$
\small\begin{equation}
{\mathbf Z}^{\dagger} = \left(
Z_\mu^C \, Z_\mu^B \, Z_\mu^A \, A_\mu
\right)
\label{eee18}
\end{equation}\normalsize
where
\small\begin{equation}
{\mathbf W} = \mathbf{R_4 R_3 R_2}\cdot {\mathbf Z},
\label{eee18b}
\end{equation}\normalsize
\small\begin{equation}
{\mathbf{\mathcal M}}^2 = (\mathbf{R_4 R_3 R_2})^{\dagger} \cdot {\mathbf M}^2 \cdot (\mathbf{R_4 R_3 R_2})
\label{eee20},
\end{equation}\normalsize
and
\small\begin{equation}
\mathbf{R_4 R_3 R_2}=\left(
\begin{array}{rrrr}
0                & 0                             &                              c{\theta_w} &                             s{\theta_w}\\
0                &                c{\theta_3} &               -s{\theta_3}s{\theta_w} &               s{\theta_3}c{\theta_w} \\
 c{\theta_4}  & -s{\theta_4}s{\theta_3} & -s{\theta_4}c{\theta_3}s{\theta_w} & s{\theta_4}c{\theta_3}c{\theta_w} \\
-s{\theta_4}  & -c{\theta_4}s{\theta_3} & -c{\theta_4}c{\theta_3}s{\theta_w} & c{\theta_4}c{\theta_3}c{\theta_w}
\end{array}
\right)
\label{eee19}
\end{equation}\normalsize
with $s{\theta} = \sin{\theta}$ and $c{\theta} = \cos{\theta}$.
We can describe the new gauge bosons in a parameterized way where just the $A_\mu$ boson is a mass eigenstate
\small\begin{eqnarray}
A_\mu &=& \frac{1}{\sqrt{1+2t^2}}\left[t~h_\mu^3 - \frac{t}{\sqrt{3}}~ h_\mu^8 + t \sqrt{\frac{2}{3}} ~h_\mu^{15} + d_\mu \right],
\nonumber\\
Z_\mu^A &=& \frac{1}{\sqrt{1+2t^2}}\left[\sqrt{1+t^2}~h_\mu^3 - \frac{t}{\sqrt{1+t^2}}\left(-\frac{t}{\sqrt{3}} ~h_\mu^8+ t \sqrt{\frac{2}{3}} h_\mu^{15} + d_\mu \right)\right],
\nonumber\\
Z_\mu^B &=& \frac{1}{\sqrt{3+3t^2}}\left[\sqrt{3+2t^2}~h_\mu^8 + \sqrt{\frac{3t^2}{3+2t^2}}\left( t \sqrt{\frac{2}{3}} h_\mu^{15} + d_\mu \right)\right],
\nonumber\\
Z_\mu^C &=& \frac{1}{\sqrt{3+2t^2}}\left[\sqrt{3}~h_\mu^{15} - \sqrt{2}~t~ d_\mu \right],
\end{eqnarray}\normalsize
and the $U(1)_Y$ and $U(1)_{Y_{SU(3)}}$ gauge bosons are respectively
\small\begin{eqnarray}
B_\mu~~~~&=& \frac{1}{\sqrt{1+t^2}}\left[ - \frac{t}{\sqrt{3}}~h_\mu^8 + t \sqrt{\frac{2}{3}} ~h_\mu^{15} + d_\mu \right],
\nonumber\\
B_\mu^{SU(3)} &=& \sqrt{\frac{3}{3+2t^2}}\left[  t \sqrt{\frac{2}{3}} ~h_\mu^{15} + d_\mu \right].
\end{eqnarray}\normalsize
In this way, the relation between the parameter $t$ and the mixing angles can be established as
\small\begin{equation}
\begin{array}{llrccccllrcc}
s{\theta_w}&= & \frac{t}{\sqrt{1+2t^2}}    & &&& & c {\theta_w} &= &\sqrt{\frac{1+t^2}{1+2t^2}}, &&\\
s{\theta_3}&= &-\frac{t}{\sqrt{3+3t^2}}    & &&& & c {\theta_3} &= &\sqrt{\frac{3+2t^2}{3+3t^2}}, &&\\
s{\theta_4}&= & \sqrt{\frac{2t^2}{3+2t^2}} & &&& & c {\theta_4} &= &\sqrt{\frac{3}{3+2t^2}},&&
\end{array}
\label{eee22}
\end{equation}\normalsize
with the electromagnetic coupling constant $e$ defined by
\small\begin{equation}
e=\frac{g~t}{\sqrt{1+2t^2}}=\frac{g^{\prime}}{\sqrt{1+2t^2}}
\label{eee23},
\end{equation}\normalsize
and the covariant derivative rewritten as
\footnotesize
\small\begin{eqnarray}
\fl D_\mu =  \partial _\mu  - i g_4\left[W^-_\mu W^- + W^+_\mu W^+ + U^0_\mu U^- +U^{0\dagger}_\mu  U^+ + V^+_\mu V^- + V^-_\mu V^+\right] \nonumber \\ 
\fl~~~~~~- i g_4\left[X^-_\mu X^- + X^+_\mu X^+ + Y^0_\mu Y^- + Y^{0\dagger}_\mu Y^+ + T^-_\mu T^- + T^+_\mu T^+\right] \nonumber \\
\fl~~~~~~- i  \left[g_{4} c{\theta_4} H_{15} - g_{Bl} s{\theta_4} \frac{B-l}{2}\right] Z^{C}_\mu -i \left[g_{4} \left(c{\theta_3}H_{8} - s{\theta_3}s{\theta_4} H_{15}\right) - g_{Bl} s{\theta_3}c{\theta_4} \frac{B-l}{2}\right]Z^{B}_\mu\nonumber\\
\fl~~~~~~- i  \left[g_{4} \left(c{\theta_w}H_{3} - s{\theta_w}s{\theta_3} H_{8}- s{\theta_w}c{\theta_3}s{\theta_4} H_{15}\right) - g_{Bl} s{\theta_w}c{\theta_3}c{\theta_4} \frac{B-l}{2}\right]Z^{A}_\mu - i  e Q A_\mu  ,
\label{eee24}
\end{eqnarray}\normalsize
\normalsize

The presence of six direct free parameters ($u$, $v$, $d$, $w$, $g_4$ and $g_{Bl}$) demands for both theoretical and 
phenomenological approximations which will be
suggested in the following section.

\section{\label{sec3}Mixing and gauge boson mass phenomenological constraints}

We can introduce a theoretical constraint between the coupling constants $g_4$ and $g_{Bl}$ if we consider the lines of the matrix representation of $\Psi$ in Eq.~(\ref{bbb3}) as a $4-dim$ multiplet of $SU(4)_{PS}$.
As a consequence of such a description we obtain
the following properties
\small\begin{equation}
\begin{array}{lclcl}
Tr[g^2_{4} H_a^2] &  &=&  & \frac{1}{2} g^2_{4}
\end{array}
\label{tr1},
\end{equation}\normalsize
applied to the normalized generators $H_a$ of $SU(4)_{PS}$ and
\small\begin{equation}
\begin{array}{lclcl}
Tr[g^2_{Bl} \left(\frac{B - l}{2}\right)^2] & = & \frac{1}{4}\left[\frac{1}{9} + \frac{1}{9} +\frac{1}{9} +1\right] &=& \frac{1}{3} g^2_{Bl}
\end{array}
\label{tr2},
\end{equation}\normalsize
related to the hypercharge of $U(1)_{B-l}$ summed over three quarks and one lepton.
Moreover, from (\ref{tr1}) and (\ref{tr2}),
the obvious relation $g_4 = g_{3L} = g$ obtained with the SSB chain in Eq.~(\ref{eee1})
allows us to derive the following relation between $g_4$ and $g_{Bl}$
\small\begin{equation}
g_4 = g_{c} = \sqrt{\frac{2}{3}}g_{Bl}.
\label{eee25}
\end{equation}\normalsize
It provides the correct $B-l$ quantum numbers for all fermions included in the model with the coupled symmetries expressed by Eq.~(\ref{eee1}).
Such a proposition makes $t=\sqrt{\frac{3}{2}}$ so that the mixing angles are obtained from
\small\begin{equation}
\begin{array}{lcrclcrclcr}
s{\theta_w}&=& \sqrt{\frac{3}{8}}     &~~~& c{\theta_w} &=& \sqrt{\frac{5}{8}}  &~~~& g^{\prime} &=&   \sqrt{\frac{3}{5}}g\\
s{\theta_3}&=&  \frac{1}{\sqrt{5}}    &~~~& c{\theta_3} &=& \frac{2}{\sqrt{5}}  &~~~& g_{Y_{SU(3)}}      &=&   \frac{1}{2} g_3       \\
s{\theta_4}&=& -\frac{\sqrt{2}}{2}    &~~~& c{\theta_4} &=& \frac{\sqrt{2}}{2}  &~~~& g_{Bl}    &=& \sqrt{\frac{3}{2}} g_4 
\end{array}
\label{422}
\end{equation}\normalsize
where we can immediately observe the correspondence with the value of the electroweak mixing angle
by the SSB of $SU(5)$ model \cite{GG74}.

In order to simplify the calculations,
we will use three different approximations over the energy parameters $u$, $v$, $d$ and $w$,
which perhaps should be determined by the most suitable phenomenological analysis, i.e. at this point,
the choice of such approximations is purely arbitrary.
In the first case we shall adopt $u = v \ll d = w$ which we believe to reflect correctly the low energy phenomenology
\footnote{The approximation adopting $u = d = v \ll w$ is not possible because the results subsequently obtained are
not related to the low energy phenomenology. They provide physically impossible values to the electroweak mixing angle.}.
In the second case we shall adopt $u \ll d = v \ll w$ isolating the $SU(4)_{EW}$ symmetry in a high energy scale.
The last case corresponds to a toy model which will be related to a crude approximation where we adopt $u = d = v = w$.
A most rigorous procedure should be guided by some numerical calculations from which the above parameters
could be obtained in order to accurately describe the experimental data that, as a first general attempt, we have described by
means of a rough estimation of the parameters.

\paragraph*{Case 1\\}

If we adopt $v = u \ll d = w$ with $t = \sqrt{\frac{3}{2}}$, we will obtain
\small\begin{equation}
{\mathbf{\mathcal M}}^2=\frac{g_4^2}{4}\left[
\begin{array}{rrrr}
 \frac{5v^2 + 13w^2}{3}        &  \frac{2v^2+ 10w^2}{3\sqrt{5}}  & \frac{2\sqrt{5}v^2}{\sqrt{6}}        & ~~~0  \\
\frac{2v^2+ 10w^2}{3\sqrt{5}} &  \frac{17v^2 + 25w^2}{15}       & -\frac{2\sqrt{5}v^2}{\sqrt{6}}        & ~~~0  \\
\frac{2\sqrt{5}v^2}{\sqrt{6}} & -\frac{2\sqrt{5}v^2}{\sqrt{6}}   & \frac{16v^2}{5}                       & ~~~0  \\
0                              &  0                              &  0                                    & ~~~0
\end{array}
\right].
\label{424}
\end{equation}\normalsize
The mass eigenstates $Z_\mu^{\prime\prime}$,$Z_\mu^{\prime}$, $Z_\mu^0$ and $A_\mu$ with corresponding eigenvalues become
\small\begin{equation}
\begin{array}{lcl}
A_\mu                  &=& A_\mu                                                                 \\
M^2_A                  &=& 0                                                                     \\
\end{array}
,\end{equation}\normalsize
\small\begin{equation}
\begin{array}{lcl}
Z_\mu^0                &\propto& {\mathbf W} . \left(\begin{array}{c} 5+\frac{-7v^2+25w^2}{\sqrt{25 v^4 - 14 v^2 w^2 +25w^4}} \\ \sqrt{3}\left(1+\frac{5v^2+5w^2}{\sqrt{25 v^4 - 14 v^2 w^2 +25w^4}}\right) \\  -\sqrt{6}\left(1+\frac{5v^2+5w^2}{\sqrt{25 v^4 - 14 v^2 w^2 +25w^4}}\right)\\ \sqrt{6}\left(-1+\frac{11v^2-5w^2}{\sqrt{25 v^4 - 14 v^2 w^2 +25w^4}}\right) \end{array}\right)\\
    &&\\
                       &\propto&    - \sqrt{5} (a^+) Z_\mu^C    + (a^+) Z_\mu^B + Z_\mu^A                        \\
                       &\approx& \frac{1}{2\sqrt{10}} \left( 5 ~h_\mu^3 + \sqrt{3}~h_\mu^8  -\sqrt{6}~ h_\mu^{15} -\sqrt{6}~ d_\mu \right)\\
&&\\
M^2_{Z^0}              &=&  \frac{1}{8}g_4^2(5 v^2 + 5w^2 -  5\sqrt{25 v^4 - 14 v^2 w^2 +25w^4}) \\
                       &\approx&  \frac{4}{5}g_4^2 v^2  \\
\end{array}
,\end{equation}\normalsize
\small\begin{equation}
\begin{array}{lcl}
Z_\mu^{\prime}         &=&  \frac{1}{\sqrt{5}}Z_\mu^C  +  Z_\mu^B\\
                       &=&  \frac{1}{\sqrt{3}}\left(\sqrt{2}~h_\mu^8  + ~h_\mu^{15}\right) \\
&&\\
M^2_{Z^{\prime}}       &=&  \frac{1}{4}g_4^2 (v^2 + w^2)                                         \\
\end{array}
\end{equation}\normalsize
and
\small\begin{equation}
\begin{array}{lcl}
Z_\mu^{\prime\prime}   &\propto& {\mathbf W} . \left(\begin{array}{c} 5-\frac{-7v^2+25w^2}{\sqrt{25 v^4 - 14 v^2 w^2 +25w^4}} \\ \sqrt{3}\left(1-\frac{5v^2+5w^2}{\sqrt{25 v^4 - 14 v^2 w^2 +25w^4}}\right) \\  -\sqrt{6}\left(1-\frac{5v^2+5w^2}{\sqrt{25 v^4 - 14 v^2 w^2 +25w^4}}\right)\\ -\sqrt{6}\left(1+\frac{11v^2-5w^2}{\sqrt{25 v^4 - 14 v^2 w^2 +25w^4}}\right) \end{array}\right)\\
    &&\\
                       &\propto&    -\sqrt{5} (a^-) Z_\mu^C + (a^-) Z_\mu^B + Z_\mu^A                      \\
                       &\approx& \frac{1}{\sqrt{15}} \left(-\sqrt{2}  ~h_\mu^8  + 2 ~ h_\mu^{15} - 3~ d_\mu \right)\\
&&\\
M^2_{Z^{\prime\prime}} &=&  \frac{1}{8}g_4^2(5 v^2 + 5w^2 +  5\sqrt{25 v^4 - 14 v^2 w^2 +25w^4}) \\
                       &\approx&  \frac{1}{4}g_4^2\left(5w^2 +\frac{9}{5} v^2 \right) \\
\end{array}
\end{equation}\normalsize
with
\small\begin{equation}
\begin{array}{lcl}
a^{\pm}&=      & \frac{7 v^2  - 25w^2 \pm  5\sqrt{25 v^4 - 14 v^2 w^2 +25w^4}}{24 \sqrt{6}v^2}\\
\end{array}
.\end{equation}\normalsize
The observed mass hierarchy is described by $M^2_{Z^{\prime\prime}} > M^2_{Z^{\prime}} > M^2_{Z^{0}}$.
The values of $M^2_{Z^{0}}$ and $M^2_{W}$ can be used for obtaining
\small\begin{equation}
\frac{M^2_{Z^{0}}}{M^2_{W}} \approx \frac{8}{5}
\label{apap1},
\end{equation}\normalsize
and if we attempt to the unification energy scale, the value of (\ref{apap1}) is in perfect agreement with
the ratio
\small\begin{equation}
\frac{M^2_{Z^{0}}}{M^2_{W}} \approx \frac{1}{\cos{\theta_w}^2}
,\label{apa1}
\end{equation}\normalsize
which gives the same value obtained from the SM without radiative corrections.
It is an important result which not only corroborates the positive effect of assuming a theoretical value $t = \sqrt{\frac{3}{2}}$ but also agrees with the value obtained in the $SU(5)$ model \cite{GG74}.
It means that the corresponding renormalization calculations can be extended to our model.

\paragraph*{Case 2\\}

If we adopt $u \ll d = v \ll w$ with $t = \sqrt{\frac{3}{2}}$ and we consider the parameter $v$ related to the GWS electroweak energy scale, we will be able to make $u \approx 0$ and obtain
\small\begin{equation}
{\mathbf{\mathcal M}}^2=\frac{g_4^2}{4}\left[
\begin{array}{rrrr}
\frac{8v^2 + w^2}{3}         & -\frac{8v^2}{3\sqrt{5}}         & \frac{4\sqrt{2}v^2}{\sqrt{15}}     & ~~~0  \\
-\frac{8v^2}{3\sqrt{5}}         &  \frac{26v^2}{5}                & \frac{2\sqrt{2}v^2}{5\sqrt{3}}       & ~~~0  \\
 \frac{4\sqrt{2}v^2}{\sqrt{15}} &  \frac{2\sqrt{2}v^2}{5\sqrt{3}} & \frac{8v^2}{5}                       & ~~~0  \\
0                               &  0                              &  0                                    & ~~~0
\end{array}
\right].
\label{4242}
\end{equation}\normalsize
The mass eigenstates $Z_\mu^{\prime\prime}$,$Z_\mu^{\prime}$, $Z_\mu^0$ and $A_\mu$ with corresponding eigenvalues become
\small\begin{equation}
\begin{array}{lcl}
A_\mu                  &=& A_\mu                                                                 \\
M^2_A                  &=& 0                                                                     \\
\end{array}
,\end{equation}\normalsize
\small\begin{equation}
\begin{array}{lcl}
Z_\mu^0                &\propto& {\mathbf W} . \left(\begin{array}{c} 1 \\-\frac{1}{\sqrt{3}}\\  \frac{(-4 v^2  + 9w^2 -  3\sqrt{16 v^4 + 8v^2 w^2 +9w^4})}{4\sqrt{6}v^2}\\ \frac{(-12 v^2  -9 w^2 + 3\sqrt{16 v^4 + 8v^2 w^2 +9w^4})}{4\sqrt{6}v^2} \end{array} \right)\\
    &&\\
                       &\propto& \sqrt{5}(b^-) Z_\mu^C  - \sqrt{\frac{2}{3}} Z_\mu^B + Z_\mu^A                       \\
                       &\approx& \frac{1}{\sqrt{13}} \left( \sqrt{3} ~h_\mu^3  - ~h_\mu^8  -\frac{3\sqrt{2}}{2}~ h_\mu^{15} -\frac{3\sqrt{2}}{2}~ d_\mu \right)\\
&&\\
M^2_{Z^0}              &=&  \frac{1}{8}g_4^2(4v^2 + 3w^2 - \sqrt{16 v^4 + 8 v^2 w^2 + 9w^4}) \\
                       &\approx&  \frac{1}{3}g_4^2 v^2  \\
\end{array}
,\end{equation}\normalsize
\small\begin{equation}
\begin{array}{lcl}
Z_\mu^{\prime}         &=&  \sqrt{\frac{2}{3}}Z_\mu^B  +  Z_\mu^A                                    \\
                       &=&  \frac{1}{2}\left(h_\mu^{3} - \sqrt{3}~h_\mu^{8}\right)\\
&&\\
M^2_{Z^{\prime}}       &=&  \frac{1}{2}g_4^2 v^2                                          \\
\end{array}
\end{equation}\normalsize
and
\small\begin{equation}
\begin{array}{lcl}
Z_\mu^{\prime\prime}   &\propto& {\mathbf W} . \left(\begin{array}{c} 1 \\-\frac{1}{\sqrt{3}}\\  \frac{(-4 v^2  + 9w^2 +  3\sqrt{16 v^4 + 8v^2 w^2 +9w^4})}{4\sqrt{6}v^2}\\ \frac{(-12 v^2  -9 w^2 - 3\sqrt{16 v^4 + 8v^2 w^2 +9w^4})}{4\sqrt{6}v^2}\end{array}\right)\\
    &&\\
                       &\propto&   \sqrt{5}(b^+) Z_\mu^C    - \sqrt{\frac{2}{3}} Z_\mu^B + Z_\mu^A                      \\
                       &\approx& \frac{1}{\sqrt{2}} \left(h_\mu^{15} -~ d_\mu \right)\\
&&\\
M^2_{Z^{\prime\prime}} &=&  \frac{1}{8}g_4^2(4v^2 + 3w^2 + \sqrt{16 v^4 + 8 v^2 w^2 + 9w^4}) \\
                       &\approx&  \frac{1}{4}g_4^2\left(3w^2 +\frac{8}{3} v^2 \right) \\
\end{array}
\end{equation}\normalsize
with
\small\begin{equation}
\begin{array}{lcl}
b^{\pm}&=      & \frac{\sqrt{5}(4 v^2  + 9w^2 \pm  3\sqrt{16 v^4 + 8v^2 w^2 +9w^4})}{8\sqrt{6}v^2}\\
\end{array}
.\end{equation}\normalsize
The observed mass hierarchy is described again by $M^2_{Z^{\prime\prime}} > M^2_{Z^{\prime}} > M^2_{Z^{0}}$ and the values of $M^2_{Z^{0}}$ and $M^2_{W}$ can be used for obtaining
\small\begin{equation}
\frac{M^2_{Z^{0}}}{M^2_{W}} \approx \frac{4}{3}
\label{apap2}.
\end{equation}\normalsize
Experimentally, at low energies, the value of (\ref{apap2}) is $1.3$,
and it is not far from a real physical situation. But it is remarkable
that the renormalization calculation were not considered.

\paragraph*{Case 3\\}

Here we apply $u = d = w = v$ trying to make some comparisons.
Firstly we obtain
\small\begin{equation}
{\mathbf{\mathcal M}}^2=\frac{g_4^2}{4}\left[
\begin{array}{rrrr}
   6v^2                        & -\frac{4v^2}{\sqrt{5}}  & \frac{2\sqrt{6}v^2}{\sqrt{5}}     & 0  \\
-\frac{4v^2}{\sqrt{5}}         &  \frac{14v^2}{5}        &-\frac{2\sqrt{6}v^2}{5}       & 0  \\
 \frac{2\sqrt{6}v^2}{\sqrt{5}} & -\frac{2\sqrt{6}v^2}{5} & \frac{16v^2}{5}                       & 0  \\
0                              &  0                      &  0                                    & 0
\end{array}
\right].
\label{4242b}
\end{equation}\normalsize
The mass eigenstates $Z_\mu^{\prime\prime}$,$Z_\mu^{\prime}$, $Z_\mu^0$ and $A_\mu$ with corresponding eigenvalues become
\small\begin{equation}
\begin{array}{lcl}
A_\mu                  &=& A_\mu                                                                 \\
M^2_A                  &=& 0                                                                     \\
\end{array}
,\end{equation}\normalsize
\small\begin{equation}
\begin{array}{lcl}
Z_\mu^0                &=&  -\sqrt{\frac{10}{3}} Z_\mu^C + Z_\mu^A                       \\
                       &=& \frac{1}{\sqrt{52}} \left(5~h_\mu^{3} + \sqrt{3}~h_\mu^{8} -2 \sqrt{6}~h_\mu^{15}\right)\\
&&\\
M^2_{Z^0}              &=&  \frac{1}{2}g_4^2 v^2  \\
\end{array}
,\end{equation}\normalsize
\small\begin{equation}
\begin{array}{lcl}
Z_\mu^{\prime}         &=&  \frac{1}{\sqrt{5}}Z_\mu^C  +  Z_\mu^A                                    \\
                       &=& \frac{1}{\sqrt{3}} \left(\sqrt{2}~h_\mu^{8} + ~h_\mu^{15}\right)\\
&&\\
M^2_{Z^{\prime}}       &=&  \frac{1}{2}g_4^2 v^2                                          \\
\end{array}
\end{equation}\normalsize
and
\small\begin{equation}
\begin{array}{lcl}
Z_\mu^{\prime\prime}   &=&  \sqrt{\frac{10}{3}}Z_\mu^C  - \sqrt{\frac{2}{3}} Z_\mu^B + Z_\mu^A                      \\
                       &=& \frac{1}{2\sqrt{6}} \left(\sqrt{3}~h_\mu^{3} - ~h_\mu^{8} + \sqrt{2}~h_\mu^{15} - 3 \sqrt{2} ~d_\mu\right)\\
&&\\
M^2_{Z^{\prime\prime}} &=&  2 g_4^2 v^2
\end{array}
\end{equation}\normalsize
with the mass hierarchy $M^2_{Z^{\prime\prime}} > M^2_{Z^{\prime}}, \,  M^2_{Z^{0}}$ and the relation
\small\begin{equation}
\frac{M^2_{Z^{0}}}{M^2_{W}} \approx 1
\label{apap3}.
\end{equation}\normalsize
The value found in (\ref{apap3}) corresponds approximately to the value found by
Pisano and Pleitez \cite{PP95} in the use of a different value for $t$ which was
chosen in order to obtain the right value for neutral coupling for quarks.
It also suggests that the $Z^{\prime}$, $Z^0$ are $W^\pm$ mass degenerate at tree level.
Hence, as observed by Pisano and Pleitez, the right value of (\ref{apap3}) must
arise only by means of radiative corrections.

\section{\label{sec4}The charged and neutral currents}

We can write the charged and neutral current Lagrangian terms obtained from (\ref{eee24})
with their no null coupling terms as 
\small\begin{eqnarray}
\mathcal{L} & = & \overline{\psi^b_{(LR)}}(i \gamma^{\mu}) D_{\mu}\psi^b_{(LR)} \nonumber\\		
			& = & \overline{\psi^b_{(LR)}}(i \gamma^{\mu}) \partial_{\mu}\psi^b_{(LR)} +  g_4\left(W^+_\mu J^{\mu}_{W^+} + W^-_\mu J^{\mu}_{W^-} + X^+_\mu J^{\mu}_{X^+} + X^-_\mu J^{\mu}_{X^-}\right) \nonumber\\ 
			&   & +  g_4 \left( Z_\mu J^{\mu}_Z + Z^{\prime}_\mu J^{\mu}_{Z^{\prime}} + Z^{\prime\prime}_\mu J^{\mu}_{Z^{\prime\prime}}\right) + e A_ \mu J^{\mu}_{EM},
\label{coc1}
\end{eqnarray}\normalsize
with $b =$ 1 (red), 2 (green), 3 (blue) and 4 ($\ell$).
If the gauge bosons $V_\mu^{\mp}$, $T_\mu^{\pm}$, $U_\mu^{0}$ and $Y_\mu^{0}$ couple
with left and right chiral states at the same time, the coupling amplitude will vanish
because $\gamma^\mu$ anticommutes with $\gamma^5$.
The remaining charged currents can be easily written as 
\small\begin{eqnarray}
J^{\mu}_{W^+}   &=&  \frac{1}{\sqrt{2}}	 	  \left(\overline{\nu_L} \gamma ^\mu e_L + \overline{u_L} \gamma ^\mu d_L\right),                                	   			 				  			   		   		 		 		     \nonumber\\
J^{\mu}_{W^-}   &=&  \frac{1}{\sqrt{2}}	 	  \left(\overline{e_L} \gamma ^\mu \nu _L +  \overline{d_L} \gamma ^\mu u_L\right),                                														 				 		     \nonumber\\
J^{\mu}_{X^+}   &=&  \frac{1}{\sqrt{2}}	 	  \left(\overline{\nu_R} \gamma ^\mu e_R + \overline{u_R} \gamma ^\mu d_R\right)  ,                              	   			 				  			   		   		 		 		     \nonumber\\
J^{\mu}_{X^-}   &=&  \frac{1}{\sqrt{2}}	 	  \left(\overline{e_R} \gamma ^\mu \nu _R +  \overline{d_R} \gamma ^\mu u_R\right).                                														 				 		     \nonumber\\
\end{eqnarray}\normalsize
Obviously, the electromagnetic current follows the same rule since it has the same structure of the GWS electromagnetic current given by
\small\begin{equation}
J^{\mu+}_{EM} =            (-1) \overline{e} \gamma ^\mu e    +  \left(\frac{2}{3}\right)\overline{u} \gamma ^\mu u  + \left(-\frac{1}{3}\right)\overline{d} \gamma ^\mu d	   	  
.\end{equation}\normalsize
The remaining neutral currents shall be written in agreement with
each of the three approximation cases analyzed in the previous section
since a general term would become a little complicated for obtaining the phenomenological constraints.

In the first case we have
\small\begin{equation}
Z_\mu^0 			   \approx \frac{1}{2\sqrt{10}} \left( 5 ~h_\mu^3 + \sqrt{3}~h_\mu^8  -\sqrt{6}~ h_\mu^{15} -\sqrt{6}~ d_\mu \right)
,\end{equation}\normalsize
with
\small\begin{equation}
\fl\begin{array}{lcclcl}
J^{\mu}_{Z}   &=& & \frac{1}{2\sqrt{10}}\left(\frac{5}{2} + \frac{1}{2} - \frac{1}{2} + \frac{\sqrt{6}}{2}\right) \overline{\nu_L} \gamma ^\mu \nu_L   &+&  \frac{1}{2\sqrt{10}}\left(-\frac{5}{2} + \frac{1}{2} - \frac{1}{2} + \frac{\sqrt{6}}{2}\right) \overline{e_L} \gamma ^\mu e_L  \\	   	  
			  & &+& \frac{1}{2\sqrt{10}}\left(\frac{5}{2} + \frac{1}{2} - \frac{1}{2} - \frac{\sqrt{6}}{6}\right) \overline{u_L} \gamma ^\mu u_L       &+&  \frac{1}{2\sqrt{10}}\left(-\frac{5}{2} + \frac{1}{2} - \frac{1}{2} - \frac{\sqrt{6}}{6}\right) \overline{d_L} \gamma ^\mu d_L  \\	   	  
			  & &+& \frac{1}{2\sqrt{10}}\left(    0       -     1       - \frac{1}{2} + \frac{\sqrt{6}}{2}\right) \overline{\nu_R} \gamma ^\mu \nu_R   &+&  \frac{1}{2\sqrt{10}}\left(     0       +     0       + \frac{3}{2} + \frac{\sqrt{6}}{2}\right) \overline{e_R} \gamma ^\mu e_R  \\	   	  
			  & &+& \frac{1}{2\sqrt{10}}\left(    0       -     1       - \frac{1}{2} - \frac{\sqrt{6}}{6}\right) \overline{u_R} \gamma ^\mu u_R       &+&  \frac{1}{2\sqrt{10}}\left(     0       +     0       + \frac{3}{2} - \frac{\sqrt{6}}{6}\right) \overline{d_R} \gamma ^\mu d_R  	   	  
,\end{array}
\label{ph1a}
\end{equation}\normalsize

\small\begin{equation}
Z_\mu^{\prime}         =  \frac{1}{\sqrt{3}}\left(\sqrt{2}~h_\mu^8  + ~h_\mu^{15}\right) 
,\end{equation}\normalsize
with
\small\begin{equation}
\fl\begin{array}{lcclcl}
J^{\mu}_{Z^{\prime}}   &=& & \frac{1}{\sqrt{3}}\left(\frac{1}{\sqrt{6}} + \frac{1}{2\sqrt{6}} \right) \overline{\nu_L} \gamma ^\mu \nu_L   &+&  \frac{1}{\sqrt{3}}\left(\frac{1}{\sqrt{6}} + \frac{1}{2\sqrt{6}}\right) \overline{e_L} \gamma ^\mu e_L  \\	   	  
			  		   & &+& \frac{1}{\sqrt{3}}\left(\frac{1}{\sqrt{6}} + \frac{1}{2\sqrt{6}} \right) \overline{u_L} \gamma ^\mu u_L       &+&  \frac{1}{\sqrt{3}}\left(\frac{1}{\sqrt{6}} + \frac{1}{2\sqrt{6}}\right) \overline{d_L} \gamma ^\mu d_L  \\	   	  
			  		   & &+& \frac{1}{\sqrt{3}}\left(-\frac{2}{\sqrt{6}} + \frac{1}{2\sqrt{6}}\right) \overline{\nu_R} \gamma ^\mu \nu_R   &+&  \frac{1}{\sqrt{3}}\left(0 - \frac{3}{2\sqrt{6}}\right) \overline{e_R} \gamma ^\mu e_R  \\	   	  
			  		   & &+& \frac{1}{\sqrt{3}}\left(-\frac{2}{\sqrt{6}} + \frac{1}{2\sqrt{6}}\right) \overline{u_R} \gamma ^\mu u_R       &+&  \frac{1}{\sqrt{3}}\left(0 - \frac{3}{2\sqrt{6}}\right) \overline{d_R} \gamma ^\mu d_R  	   	  
,\end{array}
\label{ph1b}
\end{equation}\normalsize
and
\small\begin{equation}
Z_\mu^{\prime\prime}   \approx \frac{1}{\sqrt{15}} \left(-\sqrt{2}  ~h_\mu^8  + 2 ~ h_\mu^{15} - 3~ d_\mu \right)
,\end{equation}\normalsize
with
\small\begin{equation}
\fl\begin{array}{lcclcl}
J^{\mu}_{Z^{\prime\prime}}   &=& & \frac{1}{\sqrt{15}}\left(-\frac{1}{\sqrt{6}} + \frac{1}{\sqrt{6}} + \frac{3}{2}\right) \overline{\nu_L} \gamma ^\mu \nu_L   &+&  \frac{1}{\sqrt{15}}\left(-\frac{1}{\sqrt{6}} + \frac{1}{\sqrt{6}} + \frac{3}{2}\right) \overline{e_L} \gamma ^\mu e_L  \\	   	  
			  				 & &+& \frac{1}{\sqrt{15}}\left(-\frac{1}{\sqrt{6}} + \frac{1}{\sqrt{6}} - \frac{1}{2}\right) \overline{u_L} \gamma ^\mu u_L       &+&  \frac{1}{\sqrt{15}}\left(-\frac{1}{\sqrt{6}} + \frac{1}{\sqrt{6}} - \frac{1}{2}\right) \overline{d_L} \gamma ^\mu d_L  \\	   	  
			  				 & &+& \frac{1}{\sqrt{15}}\left( \frac{2}{\sqrt{6}} + \frac{1}{\sqrt{6}} + \frac{3}{2}\right) \overline{\nu_R} \gamma ^\mu \nu_R   &+&  \frac{1}{\sqrt{15}}\left( 0 - \frac{3}{\sqrt{6}} + \frac{3}{2}\right) \overline{e_R} \gamma ^\mu e_R  \\	   	  
			  				 & &+& \frac{1}{\sqrt{15}}\left( \frac{2}{\sqrt{6}} + \frac{1}{\sqrt{6}} - \frac{1}{2}\right) \overline{u_R} \gamma ^\mu u_R       &+&  \frac{1}{\sqrt{15}}\left( 0 - \frac{3}{\sqrt{6}} - \frac{1}{2}\right) \overline{d_R} \gamma ^\mu d_R   	   	  
.
\end{array}
\label{ph1c}
\end{equation}\normalsize

In the second case we have
\small\begin{equation}
Z_\mu^0                \approx \frac{1}{\sqrt{13}} \left( \sqrt{3} ~h_\mu^3  - ~h_\mu^8  -\frac{3\sqrt{2}}{2}~ h_\mu^{15} -\frac{3\sqrt{2}}{2}~ d_\mu \right)
,\end{equation}\normalsize
with
\small\begin{equation}
\fl\begin{array}{ll}
J^{\mu}_{Z}   = & \frac{1}{\sqrt{13}}\left(\frac{\sqrt{3}}{2} - \frac{1}{2\sqrt{3}} - \frac{\sqrt{3}}{4} + \frac{3\sqrt{2}}{4}\right) \overline{\nu_L} \gamma ^\mu \nu_L   +  \frac{1}{\sqrt{13}}\left( -\frac{\sqrt{3}}{2} - \frac{1}{2\sqrt{3}} - \frac{\sqrt{3}}{4} + \frac{3\sqrt{2}}{4}\right) \overline{e_L} \gamma ^\mu e_L  \\	   	  
			  &+ \frac{1}{\sqrt{13}}\left(\frac{\sqrt{3}}{2} - \frac{1}{2\sqrt{3}} - \frac{\sqrt{3}}{4} - \frac{\sqrt{2}}{4}\right) \overline{u_L} \gamma ^\mu u_L       +  \frac{1}{\sqrt{13}}\left( -\frac{\sqrt{3}}{2} - \frac{1}{2\sqrt{3}} - \frac{\sqrt{3}}{4} - \frac{\sqrt{2}}{4}\right) \overline{d_L} \gamma ^\mu d_L  \\	   	  
			  &+ \frac{1}{\sqrt{13}}\left(0                  + \frac{1}{\sqrt{3}}  - \frac{\sqrt{3}}{4} + \frac{3\sqrt{2}}{4}\right) \overline{\nu_R} \gamma ^\mu \nu_R   + \frac{1}{\sqrt{13}}\left(0                    +  0                  + \frac{3\sqrt{3}}{4} + \frac{3\sqrt{2}}{4}\right) \overline{e_R} \gamma ^\mu e_R  \\	   	  
			  &+ \frac{1}{\sqrt{13}}\left(0                  + \frac{1}{\sqrt{3}}  - \frac{\sqrt{3}}{4} - \frac{\sqrt{2}}{4}\right) \overline{u_R} \gamma ^\mu u_R       +  \frac{1}{\sqrt{13}}\left(0                    +  0                  + \frac{3\sqrt{3}}{4} - \frac{\sqrt{2}}{4}\right) \overline{d_R} \gamma ^\mu d_R   	   	  
,\end{array}
\label{ph2a}
\end{equation}\normalsize

\small\begin{equation}
Z_\mu^{\prime}         =  \frac{1}{2}\left(h_\mu^{3} - \sqrt{3}~h_\mu^{8}\right)		                            
,\end{equation}\normalsize
with
\small\begin{equation}
\fl\begin{array}{lcclclclcl}
J^{\mu}_{Z^{\prime}}   &=& & \frac{1}{2}\left(-\frac{1}{2}-\frac{1}{2}\right) \overline{e_L} \gamma ^\mu e_L    &+&  \frac{1}{2}\left(-\frac{1}{2}-\frac{1}{2}\right) \overline{d_L} \gamma ^\mu d_L  &+& \frac{1}{2}\overline{\nu_R} \gamma ^\mu \nu_R   &+&  \frac{1}{2} \overline{u_R} \gamma ^\mu u_R  	   	  
,\end{array}
\label{ph2b}
\end{equation}\normalsize
and
\small\begin{equation}
Z_\mu^{\prime\prime}   \approx \frac{1}{\sqrt{2}} \left(h_\mu^{15} -~ d_\mu \right)
,\end{equation}\normalsize
with
\small\begin{equation}
\fl\begin{array}{lcclcl}
J^{\mu}_{Z^{\prime\prime}}   &=& & \frac{1}{\sqrt{2}} \left( \frac{1}{2\sqrt{6}} + \frac{1}{2}\right) \overline{\nu_L} \gamma ^\mu \nu_L   &+&  \frac{1}{\sqrt{2}} \left(\frac{1}{2\sqrt{6}} + \frac{1}{2} \right) \overline{e_L} \gamma ^\mu e_L  \\	   	  
			  				 & &+& \frac{1}{\sqrt{2}} \left( \frac{1}{2\sqrt{6}} - \frac{1}{6}\right) \overline{u_L} \gamma ^\mu u_L       &+&  \frac{1}{\sqrt{2}} \left(\frac{1}{2\sqrt{6}} - \frac{1}{6} \right) \overline{d_L} \gamma ^\mu d_L  \\	   	  
			  				 & &+& \frac{1}{\sqrt{2}} \left( \frac{1}{2\sqrt{6}} + \frac{1}{2}\right) \overline{\nu_R} \gamma ^\mu \nu_R   &+&  \frac{1}{\sqrt{2}} \left(-\frac{3}{2\sqrt{6}} + \frac{1}{2}\right) \overline{e_R} \gamma ^\mu e_R  \\	   	  
			  				 & &+& \frac{1}{\sqrt{2}} \left( \frac{1}{2\sqrt{6}} - \frac{1}{6}\right) \overline{u_R} \gamma ^\mu u_R       &+&  \frac{1}{\sqrt{2}} \left(-\frac{3}{2\sqrt{6}} - \frac{1}{6}\right) \overline{d_R} \gamma ^\mu d_R.  	   	  
\end{array}
\label{ph2c}
\end{equation}\normalsize

And, finally, in the third case we have
\small\begin{equation}
Z_\mu^0                = \frac{1}{\sqrt{52}} \left(5~h_\mu^{3} + \sqrt{3}~h_\mu^{8} -2 \sqrt{6}~h_\mu^{15}\right)
,\end{equation}\normalsize
with
\small\begin{equation}
\fl\begin{array}{lcclcl}
J^{\mu}_{Z}   &=&  &\frac{1}{\sqrt{52}}\left(\frac{5}{2} + \frac{1}{2} -1 \right) \overline{\nu_L} \gamma ^\mu \nu_L   &+& \frac{1}{\sqrt{52}}\left(-\frac{5}{2} + \frac{1}{2} -1 \right) \overline{e_L} \gamma ^\mu e_L    \\	   	  
			  & &+&\frac{1}{\sqrt{52}}\left(\frac{5}{2} + \frac{1}{2} -1 \right) \overline{u_L} \gamma ^\mu u_L   &+& \frac{1}{\sqrt{52}}\left(-\frac{5}{2} + \frac{1}{2} -1 \right) \overline{d_L} \gamma ^\mu d_L  \\	   	  
			  & &+&\frac{1}{\sqrt{52}}\left(0           -     1       -1 \right) \overline{\nu_R} \gamma ^\mu \nu_R   &+&\frac{1}{\sqrt{52}}\left(0            + 0 +3 \right) \overline{e_R} \gamma ^\mu e_R  \\	   	  
			  & &+&\frac{1}{\sqrt{52}}\left(0           -     1       -1 \right) \overline{u_R} \gamma ^\mu u_R       &+&\frac{1}{\sqrt{52}}\left(0            + 0 +3 \right) \overline{d_R} \gamma ^\mu d_R,  
\end{array}
\label{ph3a}
\end{equation}\normalsize

\small\begin{equation}
Z_\mu^{\prime}        = \frac{1}{\sqrt{3}} \left(\sqrt{2}~h_\mu^{8} + ~h_\mu^{15}\right)
,\end{equation}\normalsize
with
\small\begin{equation}
\fl
J^{\mu}_{Z^{\prime}}= \frac{1}{2\sqrt{2}}( \overline{\nu_L} \gamma ^\mu \nu_L   + 
\overline{e_L} \gamma ^\mu e_L  + \overline{u_L} \gamma ^\mu u_L + \overline{d_L} \gamma ^\mu d_L - \overline{\nu_R}
\gamma ^\mu \nu_R  - \overline{e_R} \gamma ^\mu e_R  - \overline{u_R} \gamma ^\mu u_R      - \overline{d_R} \gamma ^\mu d_R),
\label{ph3b}
\end{equation}\normalsize
and
\small\begin{equation}
Z_\mu^{\prime\prime}   = \frac{1}{2\sqrt{6}} \left(\sqrt{3}~h_\mu^{3} - ~h_\mu^{8} + \sqrt{2}~h_\mu^{15} - 3 \sqrt{2} ~d_\mu\right)
,\end{equation}\normalsize
with
\small\begin{equation}
\fl\begin{array}{ll}
J^{\mu}_{Z^{\prime\prime}}  =& \frac{1}{2\sqrt{6}}\left(\frac{\sqrt{3}}{2} - \frac{1}{2\sqrt{3}} + \frac{1}{2\sqrt{3}} + \frac{3\sqrt{2}}{2} \right) \overline{\nu_L} \gamma ^\mu \nu_L   +  \frac{1}{2\sqrt{6}}\left(-\frac{\sqrt{3}}{2} - \frac{1}{2\sqrt{3}} + \frac{1}{2\sqrt{3}} + \frac{3\sqrt{2}}{2} \right) \overline{e_L} \gamma ^\mu e_L  \\	   	  
			  				 & + \frac{1}{2\sqrt{6}}\left(\frac{\sqrt{3}}{2} - \frac{1}{2\sqrt{3}} + \frac{1}{2\sqrt{3}} - \frac{\sqrt{2}}{2}  \right) \overline{u_L} \gamma ^\mu u_L       +  \frac{1}{2\sqrt{6}}\left(-\frac{\sqrt{3}}{2} - \frac{1}{2\sqrt{3}} + \frac{1}{2\sqrt{3}} - \frac{ \sqrt{2}}{2} \right) \overline{d_L} \gamma ^\mu d_L  \\	   	  
			  				 & + \frac{1}{2\sqrt{6}}\left(0                  + \frac{1}{\sqrt{3}}  + \frac{1}{2\sqrt{3}} + \frac{3\sqrt{2}}{2} \right) \overline{\nu_R} \gamma ^\mu \nu_R   +  \frac{1}{2\sqrt{6}}\left(0                   +       0             - \frac{3}{2\sqrt{3}} + \frac{3\sqrt{2}}{2}\right) \overline{e_R} \gamma ^\mu e_R  \\	   	  
			  				 & + \frac{1}{2\sqrt{6}}\left(0                  + \frac{1}{\sqrt{3}}  + \frac{1}{2\sqrt{3}} - \frac{\sqrt{2}}{2}  \right) \overline{u_R} \gamma ^\mu u_R       +  \frac{1}{2\sqrt{6}}\left(0                   +       0             - \frac{3}{2\sqrt{3}} - \frac{\sqrt{2}}{2}\right) \overline{d_R} \gamma ^\mu d_R.
\end{array}
\label{ph3c}
\end{equation}\normalsize

The results obtained in this section can be extended to an analysis upon the phenomenological constraints on the leptonic transition rate derived from the coupling currents which, however, in this preliminary analysis, will be carried in a very simplified way.

\section{\label{sec5}Decay rate and cross section constraints}

Beside the gauge boson mass results, the coupling currents can be used for calculating and
interpreting other variables of phenomenological relevance for the high energy physics.
By following the approaches suggested in the last section, we shall calculate some
transition rates and cross sections which can provide interesting results to be indirectly compared with
the experimental measurements.

In general, the mixing angles of an unification model are the phenomenological parameters to be determined
at the unification energy given, in agreement with our assumption, by
$M_{Z^{\prime},Z^{\prime\prime}}$.
In the literature \cite{Quigg}, this is the standard procedure for obtaining the electroweak phenomenological constraints
 since it involves some variables like the transition rate 
($\Gamma_{(\mbox{\footnotesize gauge boson} \rightarrow \overline{f}f)}$)
and the cross section ($\sigma_{(\nu \ell \rightarrow \nu \ell)}$ or $\sigma_{(\ell^- \ell^+ \rightarrow \bar{f} f)}$)
which have the values theoretically calculated at the unification scale ($M_{Z^{\prime},Z^{\prime\prime}}$).

Although we have introduced a relevant simplification
provided by the use of $g_{PS}\equiv g_{LR} \equiv g_{Bl}$
at unification energy scale, independently of renormalization calculations,
all the coupling constants tend to the same value.
Therefore, as we commonly find in the literature,
the following results can only provide some phenomenological interpretation if
the correspondence with the very high energy physics
is given by the energy value $M_{Z^{\prime},Z^{\prime\prime}}$.

\subsection{Some decay rates at unification scale}

\hspace{1 em} The VEVs depending on $u$, $v$, $w$ and $d$
which give mass to gauge bosons are free parameters.
Since we know the characteristics of the vector boson decay,
the masses can be determined.
The process mediated by charged gauge bosons are easily obtained since we know that quiral $L-R$
oscillations are not possible by means of vector gauge interactions.
The generalized decay rate expression 
\cite{Quigg} for charged vector bosons are given by
\small\begin{equation}
\Gamma = \rho_1 \frac{G_F M^3_{Boson}}{6 \pi \sqrt{2}},
\label{equa1}
\end{equation}\normalsize
where $G_F$ is the Fermi constant and $\rho_1$ is a phenomenological parameter depending
on the class of interaction produced fermions (quarks or leptons).
The electroweak radiative corrections 
could be included in some other extensions of these calculations \cite{LP98,Lan92,Hew89}.
Anyway, by considering the relation correlated with the coupling constant $g_4$ we can write
\small\begin{equation}
\frac{G_F M^2_{W}}{\sqrt{2}}=\frac{g_4^2}{8}.
\label{equa2}
\end{equation}\normalsize
However, more interesting results can be obtained from leptonic transition rates.
where the decaying modes for neutral gauge bosons are given by
\small\begin{equation}
\Gamma = \rho_1 \frac{g^2_4 M^3_{Z's}}{48 \pi M^2_W}N_c (L^2+R^2),
\label{equa5}
\end{equation}\normalsize
where $L$ and $R$ are, respectively, the left and right-handed current amplitudes,
$N_c$ is the color factor for leptons ($N_c=1$) and
quarks ($N_c=3$)\footnote{A dependence on the Cabibbo angle $\theta_C$ appear when we study decays where the flavor symmetry is broken \cite{Quigg}. It is not considered at the unification scale.}.
In order to calculate the relevant values for first generation fermions we introduce the input parameters
\cite{Eid04}
\small\begin{equation}
\begin{array}{l}
G_F = 1.16639 \times 10^{-5} ~(\hbar c)^3~ GeV^{-2},\\
M_{W} \approx 80.4 ~ GeV^{2}/c^2,\\
M_{Z^0} \approx 91.2 ~ GeV^{2}/c^2,\\
M_{Z^{\prime}} \approx > 700 ~ GeV^{2}/c^2,\\
\end{array}
\label{dados}
\end{equation}\normalsize
With the above information, by analyzing the first case of section III,  
we can introduce, as a first approximation, the following parameterization
\small\begin{equation}
\begin{array}{l}
d^2 = w^2 \approx > (215)^2~ GeV^2,\\
M_{Z^{\prime\prime}} \approx > 1562~GeV^{2}/c^2,
\end{array}
\end{equation}\normalsize
If the above value of $M_{Z^{\prime}}$ is correct, only the first case among them which were studied
can have an adequate physical correspondence.
Table \ref{ABCD}
illustrates the neutral gauge boson fermionic decay rates
for the three cases presented in section III.
Beside the input value of $M_{Z^0}$, by considering the second case, we would have
\small\begin{equation}
\begin{array}{l}
M_{Z^{\prime}} \approx 116.6 ~ GeV^{2}/c^2,\\
M_{Z^{\prime\prime}} \approx > 1562~GeV^{2}/c^2,
\end{array}
\label{dados2}
\end{equation}\normalsize
and, by considering the third case, we would have
\small\begin{equation}
\begin{array}{l}
M_{Z^{\prime}} = M_{Z^0},\\
M_{Z^{\prime\prime}} = 2 M_{Z^0},
\end{array}
\label{dados3}
\end{equation}\normalsize
\footnotesize
\begin{table}
\begin{center}
\footnotesize
\caption{\label{ABCD} Fermionic decay rates for neutral gauge bosons (first generation fermions).}  
\begin{tabular}{|l|rrrr|rrrr|rrrr|} 
\hline
                               & \multicolumn{4}{c|}{Case 1}         & \multicolumn{4}{c|}{Case 2}         & \multicolumn{4}{c|}{Case 3}\\
\cline{2-13}
  $\Gamma_{(MeV)}$                                        & $L^2$ & $R^2$ & $\Gamma$ & $\%$ & $L^2$ & $R^2$ & $\Gamma$ & $\%$ & $L^2$ & $R^2$ & $\Gamma$ & $\%$ \\
\hline
$ _{Z^0 \rightarrow \bar{\nu}\nu}$& $0.344$ & $0.002$ & $115$ & $18.4$ & $0.110$ & $0.110$ & $ 73$ & $12.6$ & $0.077$ & $0.077$ & $ 51$ & $ 7.7$ \\
$ _{Z^0 \rightarrow e^+e^-}$      & $0.040$ & $0.184$ & $ 74$ & $11.8$ & $0.021$ & $0.423$ & $147$ & $25.5$ & $0.173$ & $0.173$ & $115$ & $17.3$ \\
$ _{Z^0 \rightarrow \bar{u}u}$    & $0.108$ & $0.090$ & $197$ & $31.6$ & $0.003$ & $0.003$ & $  6$ & $ 1.0$ & $0.077$ & $0.077$ & $153$ & $23.0$ \\
$ _{Z^0 \rightarrow \bar{d}d}$    & $0.209$ & $0.030$ & $237$ & $38.0$ & $0.286$ & $0.067$ & $351$ & $60.8$ & $0.173$ & $0.173$ & $345$ & $51.9$ \\
\hline
$ _{Z^{\prime} \rightarrow \bar{\nu}\nu}$& $0.125$ & $0.125$ & $3~ 10^4$      & $12.5$ & $0.500$ & $0.000$ & $341$ & $12.5$ & $0.125$ & $0.125$ & $3~ 10^4$ & $12.5$ \\
$ _{Z^{\prime} \rightarrow e^+e^-}$      & $0.125$ & $0.125$ & $3~ 10^4$      & $12.5$ & $0.000$ & $0.500$ & $341$ & $12.5$ & $0.125$ & $0.125$ & $3~ 10^4$ & $12.5$ \\
$ _{Z^{\prime} \rightarrow \bar{u}u}$    & $0.125$ & $0.125$ & $         10^5$ & $37.5$ & $0.500$ & $0.000$ & $10^3$ & $37.5$ & $0.125$ & $0.125$ & $         10^5$ & $37.5$ \\
$ _{Z^{\prime} \rightarrow \bar{d}d}$    & $0.125$ & $0.125$ & $         10^5$ & $37.5$ & $0.000$ & $0.500$ & $10^3$ & $37.5$ & $0.125$ & $0.125$ & $         10^5$ & $37.5$ \\
\hline
$ _{Z^{\prime\prime} \rightarrow \bar{\nu}\nu}$& $0.148$ & $0.495$ & $         10^6$ & $23.3$ & $0.248$ & $0.248$ & $164$ & $29.7$ & $0.370$ & $0.370$ & $2~ 10^3$ & $37.3$ \\
$ _{Z^{\prime\prime} \rightarrow e^+e^-}$      & $0.148$ & $0.005$ & $     3~ 10^5$ & $7.0 $ & $0.248$ & $0.006$ & $ 84$ & $15.2$ & $0.065$ & $0.065$ & $345     $ & $ 6.4$ \\
$ _{Z^{\prime\prime} \rightarrow \bar{u}u}$    & $0.164$ & $0.035$ & $         10^6$ & $23.3$ & $0.001$ & $0.001$ & $  2$ & $ 0.3$ & $0.001$ & $0.001$ & $15      $ & $ 0.3$ \\
$ _{Z^{\prime\prime} \rightarrow \bar{d}d}$    & $0.164$ & $0.195$ & $     2~ 10^6$ & $46.5$ & $0.001$ & $0.303$ & $302$ & $54.7$ & $0.102$ & $0.102$ & $3~ 10^3$ & $56.0$ \\
\hline
\end{tabular}
\end{center}
\end{table}
\normalsize
where, by convenience, we have considered only three decimal units.
If we assume that flavor symmetry exists at the unification energy scale,
a correspondence with the values in table \ref{ABCD} can by established.
The values of $L$ and $R$ corresponds to the coefficients of ``left'' and ``right'' current calculated in the previous
section which couple with the neutral gauge bosons obtained for each one of the three cases analyzed.
Under these conditions, the above values for $\Gamma _{(Z^0 \rightarrow \bar{f}f)}$ can be multiplied by a factor $3$
which represents the number of generations, and thus we can notice the comparability with the experimental measurements \cite{Eid04}
in table \ref{BCD}. 
\footnotesize\begin{table}
\begin{center}
\footnotesize
\caption{\label{BCD} Correspondence with experimental values of $\Gamma _{(Z^0 \rightarrow \bar{f}f)}$.}  
\begin{tabular}{|l|rr|rr|} 
\hline
$\Gamma$                               & \multicolumn{2}{c|}{$SU(4)$}         & \multicolumn{2}{c|}{Exp.}\\
\hline
                                      & $\Gamma (MeV)$ & $\% $                   & $\Gamma (MeV)$ & $\%$        \\
$\Gamma _{(Z^0 \rightarrow \mbox{\footnotesize invisible})}$  & $345~$ & $18.4$ & $499~$  & $ 20.0$ \\
$\Gamma _{(Z^0 \rightarrow \bar{\ell}\ell)}   $               & $222~$ & $11.8$ & $250~ $ &$  10.1$\\
$\Gamma _{(Z^0 \rightarrow \mbox{\footnotesize hadrons})} $   & $1302~$& $69.6$ & $1745~$ & $ 69.9$\\
$\Gamma _{(Z^0 \rightarrow \mbox{\footnotesize Total})}$      & $1869~$& $100.0$& $2495~$ &$ 100.0$\\
\hline
\end{tabular}
\end{center}
\end{table}
\normalsize

The results presented above are just useful for comparing the three cases of approximations we have suggested.
Since we have adopted the usual electroweak scale of unification $v =  u \approx 215 \,GeV^{2}/c^{2}$ as an input parameter,
it is natural to obtain some divergent results in comparison with the GWS model predictions and the experimental statements.
It is easy to observe that implementing some corrections to the relation between $u$ and $v$,
and following some simple numerical analyisis, which corresponds to changing the electroweak scale of unification $u$ and/or $v$,
the results can be reconfigured to reproduce the experimental results here reported; (Technically, the final results depend on the choice of $u$ and $v$.
To avoid introducing a new energy subscale, we have chosen to proceed with the most simplified analysis which, obviously, can be improved).
However, because the manifestations of weak interactions become significant only at high energies, before
looking for elements of a more extended or specific calculation, the rough prediction that we have obtained
is sufficient for the guidance of some subsequent phenomenological analysis, which is out of planning for
the initial purpose of this manuscript.

\subsection{Some cross sections at unification energy scale}

\hspace{1 em} The scattering process which involves neutrinos mediated by neutral currents
are propitious for a phenomenological analysis since, from theoretical point of view,
such process are free of complications which comes with QCD.
In the models where right handed neutrinos are not considered,
the ratio between cross sections is simply given by 
\cite{PT93,LP98,Quigg}:
\small\begin{equation}
R = \frac{\sigma_{(\nu_{\mu} e \rightarrow \nu_{\mu} e)}}{\sigma_{(\overline{\nu}_{\mu} e \rightarrow \overline{\nu}_{\mu} e)}} = \frac{3 L_e^2 + R_e^2}{L_e^2 + 3 R_e^2},
\label{equa6}
\end{equation}\normalsize
However, when the right-handed neutrinos are taken into account, the coefficient $R$ is not null since now it depends on the 
left and right-handed couplings in the interaction Lagrangian.
The electroweak process involving hadrons can also be analyzed when we consider
electron-positron pair annihilation into hadrons,
\small\begin{equation}
\begin{array}{lll}
e^+ e^- &\Longrightarrow &\mu^+ \mu^-,\\
e^+ e^- &\Longrightarrow &\overline{\nu} \nu,\\
e^+ e^- &\Longrightarrow &\overline{q} q \Longrightarrow \mbox{hadrons}.
\end{array}
\end{equation}\normalsize
By assuming an annihilation process through $Z^0$, without kinematic suppression or any radiative correction,
the cross section peak for each fermion class is given by
\small\begin{equation}
\sigma^{peak}_{(e^+ e^- \rightarrow \overline{f} f)} = \frac{G_F^2 M_Z^4}{6 \pi \Gamma_Z^2} N_c(L_e^2 + R_e^2)(L_f^2 + R_f^2),
\label{equa7}
\end{equation}\normalsize
With this expression we can write the ratio at the intermediate boson peak as \cite{Quigg}:
\small\begin{equation}
R^{peak}_{(e^+ e^- \rightarrow \overline{f} f)} \equiv
 \frac{\sigma^{peak}_{(e^+ e^- \rightarrow \overline{f} f)}}{\sigma^{QED}_{(e^+ e^- \rightarrow \mu^+ \mu^-)}}
 \equiv \frac{G_F^2 M_Z^6}{32 \pi^2 \alpha^2 \Gamma_Z^2} N_c(L_e^2 + R_e^2)(L_f^2 + R_f^2),
\label{equa8}
\end{equation}\normalsize
which has to be multiplied by a factor $1/3$ when the fermion flavor symmetry is assumed.
We are adopting $\Gamma_Z =  1869 ~ MeV$ from table \ref{BCD} and,
obviously, we are inclined to take into account the values of Case 1 in table \ref{ABCD} since
it reproduces the results of the GWS model as we illustrate in table \ref{CD}.
\footnotesize
\begin{table}
\begin{center}
\footnotesize
\caption{\label{CD} Comparison between $R^{peak}_{(e^+ e^- \rightarrow \overline{f} f)}$ values of $SU(4)$ and $SU(2)_{L}$ (GWS).}  
\begin{tabular}{|l|rr|r|r|} 
\hline
       $R^{peak}$                      		  	      & $L^2$   & $R^2$    &$R_{SU(4)}$ & $R_{SU(2)}$  \\
\hline
$ R^{peak}_{(e^+ e^- \rightarrow \overline{\nu}  \nu)}$& $0.344$ & $0.002$  &$310$  & $305$ \\
$ R^{peak}_{(e^+ e^- \rightarrow \overline{\ell}\ell)}$& $0.040$ & $0.184$  &$201$  & $159$ \\
$ R^{peak}_{(e^+ e^- \rightarrow \overline{q} q^{(+\frac{2}{3})})}
  				 	 			 			  	      $& $0.108$ & $0.090$  &$530$  & $557$ \\
$ R^{peak}_{(e^+ e^- \rightarrow \overline{q} q^{(-\frac{1}{3})})}
  				 	 			 			  	      $& $0.209$ & $0.030$  &$642$  & $704$ \\
\hline
\end{tabular}
\end{center}
\end{table}
\normalsize
In spite of a relative ``distance'' from the energy scale of the experimental processes,
all the above results can serve as a guide for subsequent phenomenological studies
where higher order corrections are taken into account.
We emphasize that, for the most concordant approximation which comes from the first case,
the supposition of $v = u \ll d = w$ with $t = \sqrt{\frac{3}{2}}$ is determinant for the coincident results.
Maybe, a more generic numerical treatment for determining free parameters $u$, $v$, $d$, $w$ and $t$
followed by a phenomenological analysis could provide more accurate results. 

\section{\label{sec}Conclusion}

The idea of looking for new symmetries to describe the electroweak
unification properties has come with the ``left-right'' 221 model suggested by Mohapatra and Pati \cite{MP75}
as well as with the more recent 331 model suggested by Pisano, Pleitez and Frampton \cite{PP95,F92}
and the 442 model suggested by Foot and Filewood \cite{FF99}.
In this paper, we have introduced a discussion of some
SSB possibilities to the gauge $SU(4)$ in a similar context of the quoted models.
The main points of physical relevance obtained in the evolution of the proposed model could be summarized by
the following ones:

$i)$ In the Higgs sector, we have four $4-dim$ $SU(4)$ multiplets used in the gauge boson mass generation by means of 
a SSB. In the same way, there are other possibilities to the sequence of SSB in the use of
the $SU(4)$ adjoint representation $15-dim$ Higgs bosons.
In this case, a smaller number of Higgs multiplets would be needed for allowing fermions gaining masses.
By means of any SSB chain, in the low energy limit, the $SU(2)_L \otimes U(1)_Y$ symmetry is recovered.

$ii)$ The recovered chiral symmetry makes the model anomaly free for each particle generation.
In the electroweak scenario, the requirement from experiment that the weak interaction current are left-handed forced us to choose a chiral gauge coupling and, consequently check that the anomalous terms from the triangle diagrams cancel.
The anomalous term of a triangle diagram of three gauge bosons (a,b,c) is proportional the group theoretic invariant trace
$Tr[\gamma_{5} t^a {t^b,t^c}]$.
The presence of the chiral projector operator $\gamma_5$ in the gauge anomalous term sets the gauge anomaly cancellation for
a theory with left-right symmetry.
Our model concerns about the existence of a right-handed neutrino as a fundamental lepton so that
the discussion about anomaly cancellation is unnecessary since, for each fermion generation,
a left-right symmetry is assumed.
The same is true when we refer to mixed anomalies.
If we consider the effects of gravity on the electroweak interaction gauge theory, there is also a possibly anomalous diagram with one weak interaction and two gravitons which also is canceled in case of a left-right symmetry.

$iii)$ Despite not corresponding to a priority point, in this first analysis, discussing the mass scale for neutrinos could also be a pertinent subject.
The mass scale for neutrinos depends essentially on the mechanism we are considering for generating the effective mass value (see-saw type-I, II or III for instance)
for those particles, i.e. it depends on the number of Higgs fields and on the way that neutrinos couple with them. 
The answer to the above question is quite dependent on particular characteristics which were not discussed in this minimal version of our model.
While one can say that there exist many models which fit the observations, none (except a few) are completely predictive
and almost always they need to invoke new symmetries or new assumptions \cite{Moh04}.
By following an analogy with which is discussed in \cite{Moh04},
we can list several ways for determining the neutrino mass-scale by incorporating the right-handed neutrino (left-right symmetry) in the (double) see-saw mechanism.
The conventional see-saw mechanism requires rather high mass for the right-handed neutrino and therefore a correspondingly high scale for B - l symmetry breaking.
There is however no way at present to know what the scale of $B - l$ symmetry breaking is (in the phenomenological calculations we have supposed $w \sim 215 GeV$, which is not mandatory).
There are, for instance model bases on string compactification, where the $B - l$ is quite possibly in the $TeV$ range.
In this case small neutrino mass can be implemented by a double see-saw mechanism \cite{Moh86}.
The idea is to take a right-handed neutrino N and a singlet neutrino S which has extra quantum numbers which prevent it from coupling with the left-handed neutrino.
The results in this framework can show that, depending on the mass matrix which is proposed,
the important thing for us is that a $10\, TeV$ $B - l$ scale is enough to give neutrino masses in the $eV$ range.
Again appointed by Mohapatra, in another a class of see-saw models based on the $SO(10)$ group that embodies the left-right symmetric unification model
or the $SU(4)_c$, the mass the tau neutrino mass can be estimated provided one assumes the normal mass hierarchy for
neutrinos and a certain parameter accompanying a higher-dimensional operator to be of order 1.
The right-handed neutrino mass in a model with $16-dim$ Higgs boson arises from a non-renormalizable operator
in the scale of $100 \, GeV$, from which, by applying the see-saw mechanism, it is possible to obtain $m_{\nu_{\tau}} \approx 0.025 \,eV$,
which is close to the presently preferred value of $0.05 \, eV$. The situation with respect to other neutrino masses is however less certain and here one would have to make assumptions.

$iv)$ The situation with respect to mixing angles is much more complicated.
For instance, the striking difference between the quark and neutrino mixing angles makes one doubt whether complete quark lepton unification is truly obeyed in nature.
For the cases of normal and inverted hierarchy, we have no information on the mass of the lightest neutrino so that
we could assume it in principle to be quite small.
In that case, the general purpose with
the type-I see-saw formula along with the assumption of a diagonal Dirac neutrino mass matrix
to obtain the right-handed neutrino mass matrix is represented by $M_{R,ij} = m_{D,i}\, \mu^{-1}_{ij}\,m_{D,j}$
enables us to conclude that quite probably one of the three right-handed neutrinos is much heavier than the other two (generations).
The situation is of course completely different for the degenerate case.
This kind of separation of the RH neutrino spectrum is very suggestive of a symmetry.


$v)$ By reminding that a theoretical relation between the coupling constants was applied
to determine the electroweak mixing angle, we have reproduced the same result
obtained from the $SU(5)$ model \cite{GG74} in the high energy limit.
The minimal version of the model here developed also survives the proton decay problem since which the natural choice of
the fermion multiplets here adopted, the proton decay is prohibited in this unification scale, which is compatible
with the no observation of its decays at the present experimental level.
By embedding the $SU(4)_{PS} \otimes SU(4)_{EW}$ in a higher symmetry (for instance, when $G$ of the Eq~(\ref{bbb3b}) is the $SU(16)$) the proton decay may be reconsidered at much higher energy scale.

Just to conclude, with respect to the evaluation of the calculations, we have adopted
some simple approximations rather than a generic numeric calculation \cite{PP95,LP98}. 
With such simple approximations, the gauge boson masses as well as the current coupling
were easily obtained and could be used in some extensions of the model
to determine the leptonic transition rates and some other phenomenological parameters.
Obviously, by using some numerical methods, it would be possible to find a set of relations among
free parameters by trying to minimize the errors related to the phenomenological variables.
We have noticed that there are several remaining ways to extend the obtained results
in order to verify more accurately some phenomenological constraints, in particular,
those ones related to the leptonic transition rates and the cross sections in the scattering processes
at the real experimental energy scale.

\subsection*{Acknowledgments}

The author thanks FAPESP (PD 04/13770-0) for Financial Support.

\end{document}